\documentstyle[
	aps,
	twocolumn, 
	twocolumn
	]{revtex}

\begin{document}

\title{Power-Law Distributions and L\'evy-Stable
Intermittent Fluctuations in 
Stochastic Systems of Many
Autocatalytic Elements}
\author{Ofer Malcai,
Ofer Biham
and
Sorin Solomon}
\address{Racah Institute of Physics,
The Hebrew University,
Jerusalem 91904,Israel}
\maketitle

\begin{abstract}

A generic model of stochastic autocatalytic 
dynamics with many degrees
of freedom
$w_i$, 
$i=1,\dots,N$
is studied using computer 
simulations.
The time evolution of the $w_i$'s
combines a random multiplicative
dynamics
$w_i(t+1) = \lambda w_i(t)$
at the
individual level with a global coupling 
through a constraint
which does not allow 
the $w_i$'s
to fall below a lower 
cutoff given by
$c \cdot \bar w$,
where $\bar w$ is their momentary average and 
$0<c<1$
is a constant. 
The dynamic variables $w_i$ are found to exhibit a power-law
distribution of the form $p(w) \sim w^{-1-\alpha}$.
The exponent $\alpha (c,N)$ is quite
insensitive to the distribution $\Pi(\lambda)$
of the random factor $\lambda$, but it
is non-universal, and increases monotonically as a function of $c$.
The "thermodynamic"
limit $N \rightarrow \infty$ 
and the limit of decoupled free multiplicative random 
walks $c \rightarrow 0$
do not commute:
$\alpha(0,N) = 0$ for any finite $N$
while
$ \alpha(c,\infty) \ge 1$
(which is the common range in empirical systems)
for  any positive $c$.
The time evolution of ${\bar w (t)} $
exhibits intermittent fluctuations parametrized
by a (truncated) L\'evy-stable distribution 
$L_{\alpha}( r )$ with the same index $\alpha$.
This non-trivial relation between the 
distribution of the $w_i$'s at a given time and
the temporal fluctuations of their average 
is examined and its relevance to empirical 
systems is discussed. 
\end{abstract} 

\section{Introduction}

The origins of
power-law distributions
as well as their 
conceptual implications have been an active topic of research
in recent years.
Power laws are intrinsically related to the emergence of macroscopic
features which are scale invariant within some bounds and distinct from the
microscopic elementary degrees of freedom. 
Often, these features are insensitive 
to the details of the microscopic structures. 
Well known examples of power law 
distributions include the energy distribution between scales in turbulence
\cite{Kolmogorov62},
the distribution of earthquake magnitudes\cite{Gutenberg56},
the diameter distribution of craters and asteroids\cite{Mizutani80},
the distribution of
 city populations \cite{Zipf49,Zanette97},
the distributions of income and of wealth
\cite{Pareto97,Mandelbrot61,Mandelbrot51,Mandelbrot63,Atkinson78,Takayasu97,Anderson97},
the size-distribution of business firms
\cite{Simon58,Simon77}
and the distribution 
of the frequency of appearance of words in texts\cite{Zipf49}.
The fact that 
multiplicative dynamics tends to generate 
power-law distributions was intuitively 
invoked long ago
\cite{Simon58,Yule24,Champernown53,Kesten73}
but the limitations in 
computer simulation power kept the models under the constraints imposed
by the applicability 
of analytical treatment.
More recently, a broader class of models has been studied
combining computer simulations with theoretical analysis within the
Microscopic Representation paradigm proposed in Ref.
\cite{Solomon95}.
In particular, it was shown
\cite{Levy96b,Solomon96,Biham98}
that power laws appear 
in a variety of dynamical processes
and are maintained 
even under highly non-stationary conditions. 

In this paper we consider a generic model of stochastic dynamics with 
many degrees of freedom $w_i(t)$, $i=1,\dots,N$. 
The time evolution of the $w_i$'s is 
described by an asynchronous update mechanism 
in which at each time step one variable 
is chosen randomly and is multiplied by a factor 
$\lambda$ taken from a predefined 
distribution. 
In addition, there is a global coupling constraint which does not allow 
the $w_i$'s to fall below the lower cutoff given by $c \cdot \bar w$, where 
$\bar w$ is the momentary average of the $w_i$'s and $0<c<1$
is a constant. The dynamic variables $w_i$ are 
found to exhibit a power-law 
distribution of the form $p(w) \sim w^{-1-\alpha}$.
The exponent $\alpha$ is found to be 
insensitive to the distribution $\Pi(\lambda)$
of the random factor $\lambda$. However, $\alpha$ is non-universal, 
and increases 
monotonically as a function of $c$. 
In the limit $c = 0$ (where the  $w_i$'s become decoupled) 
$\alpha = 0$ for any {\it finite} $N$.
However, in the "thermodynamic" limit $N = \infty$, 
$\alpha \ge 1$ for any {\it positive} $c$. 
Thus the two limits do not commute. 
This is important for applications since typically in empirical systems
$\alpha \ge 1$ 
\cite{Kolmogorov62,Gutenberg56,Mizutani80,Zipf49,Zanette97,Pareto97,Mandelbrot61,Mandelbrot51,Mandelbrot63,Atkinson78,Takayasu97,Anderson97,Simon58,Simon77}
unlike the case of
the free
multiplicative random walk which predicts 
a log-normal distribution \cite{Gnedenko54,Aitcheson57}
corresponding to $\alpha = 0$ \cite{Montroll82}.

The time evolution of $\bar w(t)$
exhibits intermittent fluctuations parametrized by 
a truncated L\'evy-stable distribution with the same index $\alpha$. 
This intricate 
relation between the distribution of the $w_i$'s at a given 
time and the temporal 
fluctuations of their average is examined and its 
relevance to empirical systems is 
discussed. 
Our model indicates that in certain cases the scaling 
exponent may be insensitive to the 
distribution of the multiplicative (random) factor 
$\lambda$ and depends only on the "lower
bound" features which control the smallest values of the elementary variables.
The relation between the limiting conditions and the power law exponent is to be 
applied in each particular case and it constitutes a strong instrument 
in identifying 
and validating the relevant degrees of freedom responsible for the emergence of 
scaling. 

The present paper proposes to consolidate by numerical simulations the
control one has on a specific model and help in this way its further application 
to additional systems. The paper is organized as follows. In Sec. II we present 
the model. Simulations and results are reported in Sec. III, followed by a 
discussion in Sec. IV and a summary in Sec. V.

\section{The Model}

\subsection{Formal Definition}

The model\cite{Levy96b,Solomon96}
describes the evolution in discrete time of $N$ dynamic variables $w_i (t)$, 
$i=1,\dots,N$. At each time step $t$, 
an integer $i$ is chosen randomly in the range
$1 \leq i \leq N$, which is the index of the dynamic variable $w_i$
to be updated at
that time step. A random multiplicative factor 
$\lambda(t)$ is then drawn from a
given distribution $\Pi(\lambda)$, which is independent 
of $i$ and $t$ and satisfies
$\int_{\lambda} \Pi(\lambda) d{\lambda} = 1$.
This can be, for example, a uniform 
distribution in the range
$\lambda_{min} \leq \lambda \leq \lambda_{max}$,
where
$\lambda_{min}$
and
$\lambda_{max}$
are predefined limits. 
The system is then updated 
according to the following stochastic time evolution equation
\begin{eqnarray}
w_i (t+1) &=&   \lambda (t)  w_i (t)  \nonumber \\w_j (t+1) &=& w_j (t), 
\ \ \ \ \ \ j=1,\dots,N; \ j \ne i.
\label{eq:mult}
\end{eqnarray}
This is an 
asynchronous update mechanism. 
The average value of the system components at time t 
is given by

\begin{equation}
\bar w(t) = {1 \over N} \sum_{i=1}^N w_i(t).
\end{equation}
The term on the right hand side of 
Eq.~(\ref{eq:mult}) 
describes the effect of 
auto-catalysis at the individual level. 
In addition to the update rule of 
Eq.~(\ref{eq:mult}),
the value of 
the updated variable $w_i (t+1)$ is constrained to be larger or equal to 
some lower bound which is proportional to 
the momentary average value of the $w_i$'s
according to
\begin{equation}
w_i (t+1) \ge c \cdot \bar w(t)
\end{equation}
\noindent
where $0 \le c < 1$ is a constant factor. 
This constraint is imposed immediately after step
(\ref{eq:mult})
by setting
\begin{equation}
w_i (t+1) \rightarrow \max \{ w_i(t+1), c \cdot \bar w(t) \},
 \ \ \ 
\label{eq:subsidize}
\end{equation}
\noindent
where 
$\bar w(t)$,
evaluated just before the application of 
Eq.~(\ref{eq:mult}),
is used. 
This constraint describes the effect of 
auto-catalysis at the community level.

\subsection{Main Features }

Our model is characterized by a fixed (conserved) 
number of dynamic variables $N$, while the sum of their values  
is not conserved.
The conservation of the number of dynamic variables, 
which is enforced through 
the lower cutoff constraint is essential 
since otherwise the system dwindles 
over time. 
The non-conservation of   
the sum of the values of the
dynamic variables
is important as well. 
It allows to perform the multiplicative updating 
on a single variable at a time with 
no explicit binary interactions since a gain in 
$w_i$ does not require a corresponding 
immediate loss by other $w_j$'s. 
In fact, the interactions between the dynamic 
variables are implied only in the step of 
Eq.~(\ref{eq:subsidize})
in which the lower 
cutoff is imposed. 
The dynamic rule
(\ref{eq:mult})
can be described by a master 
equation for the probability distribution $p(w)$
of the form
\begin{equation}
p(w,t+1) - p(w,t) ={1 \over N}   \left[ \int_{\lambda} \Pi(\lambda)
 p(w/{\lambda},t) d{\lambda}    - p(w,t)  \right],
\label{eq:master}
\end{equation}
\noindent
where the $1/N$ factor takes into account the fact that only one  
of the $w_i$'s is
updated in each time step.
This description applies for the bulk of the distribution of the $w_i$'s
but not in the vicinity of the lower cutoff where the step of 
Eq.~(\ref{eq:subsidize})
which is not taken into account by 
Eq.~(\ref{eq:master})
may be dominant. 

For the following analysis it is convenient to normalize the 
$w_j$'s according to
\begin{equation}
w_j(t) \rightarrow w_j(t) / \bar w(t), 
\ \ \ j=1,\dots,N.
\label{eq:normalize}
\end{equation}
\noindent
As a result, the new average
$\bar w(t)$ is normalized to
\begin{equation}
\bar w (t) =
\int_{c}^N w p(w,t) dw = 1,
\label{eq:mean}
\end{equation}
\noindent
while $\sum_i w_i(t) = N \bar w = N$. 
Performing this normalization step after each 
iteration removes the non stationary part of the distribution
and amounts statistically to an overall 
multiplicative factor. 
This (time dependent) factor which represents a global 
inflation rate can be recorded at each step. 
It is convenient to represent the 
dynamics (\ref{eq:master})
on the logarithmic scale. 
In terms of the new variables
\begin{equation}
W_i  = \ln w_i,
\label{eq:logvar}
\end{equation}
Eq.~(\ref{eq:mult})
defines a random walk with
 steps of random size $\ln \lambda$:
\begin{equation}W_i(t+1) = W_i (t) + \ln \lambda.
\label{eq:random}
\end{equation}
\noindent
 The corresponding probability distribution
 $P(W)$ becomes 
\begin{equation}{ P} (W) = 
e^W p (e^{W}).
\label{eq:transf}
\end{equation}
In terms of ${ P}$ and $W$,
the master equation (\ref{eq:master}) becomes:
\begin{eqnarray}
\lefteqn{{ P}(W,t+1) - { P}(W,t) = } \nonumber \\
 & & {1 \over N} \left[ \int_{\lambda} \Pi ({\lambda}) 
{ P}(W - {\ln \lambda},t) d{\lambda} 
- { P} (W,t) \right].
\label{eq:addmast}
\end{eqnarray}
\noindent
The asymptotic stationary solution, is 
found to be \cite{Levy96b}
\begin{equation}{ P}(W) \sim e^{-\alpha W}.
\label{eq:bolt}
\end{equation}
\noindent
In terms of the original variable $w_i$,
 we get according to 
Eq.~(\ref{eq:transf}) 
a power law distribution:
\begin{equation}
p (w) = K \cdot  w^{-1-\alpha}.
\label{eq:power}
\end{equation}
\noindent
The value of the exponent $\alpha$ is determined by 
the normalization condition
[Eq.~(\ref{eq:mean})] 
divided by the probability normalization condition  
$ \int_{c}^N  p (w,t) dw = 1 $  
(in order to eliminate the constant factor $K$),
which yields:
\begin{equation}
N =  {{\alpha-1} \over { \alpha}} 
\left[ {{ \left(c \over N \right)}^{\alpha} - 1}
\over {{ \left(c \over N \right)}^{\alpha} - 
\left(c \over N \right)} \right].
\label{eq:calc}
\end{equation}
\noindent
The exponent $\alpha$ is given implicitly 
as a function of $c$ and $N$
by  
Eq.~(\ref{eq:calc}).
We identify two regimes within $0 \le c < 1$ in which 
Eq.~(\ref{eq:calc})
can be simplified and $\alpha$ can be obtained explicitly.  
For a given $N$
and values of  $c$ in the range 
$1/\ln N \ll c < 1$  
one obtains $\alpha > 1$ as well as 
$(c/N)^{\alpha} \ll c/N \ll 1$.
Consequently, in this range, 
one can neglect 
the 
$(c/N)^{\alpha}$ 
terms in 
Eq.~(\ref{eq:calc}) 
to obtain to a good approximation
\begin{equation}
N =  {{\alpha-1} \over { \alpha}}\left[ { { - 1} \over 
{  - \left(c \over N \right)}  } \right].
\label{eq:calc1}
\end{equation}
which gives the explicit, $N$-independent solution
\begin{equation}
\alpha \cong {1 \over {1- c}}.
\label{eq:temp}
\end{equation}
\noindent
This relation is exact in the "thermodynamic" limit $N = \infty$.
The relation 
(\ref{eq:temp})  
has two remarkable properties:
(a) it does not depend
on the distribution $\Pi (\lambda )$;
(b) it gives rise to $\alpha$ values in the experimentally 
realistic range $\alpha \ge 1$.

For finite $N$ and values of $c$ lower 
than $1/ \ln N$ the approximation 
Eq.~(\ref{eq:temp})  
breaks down and values $\alpha <1$ become possible.
However, for any finite N, 
another approximation holds in the range
$c \ll 1/N < 1$.
In this range
$(c/N) \ll (c/N)^{\alpha} \ll 1$ 
and therefore one can neglect $ (c/N)^{\alpha}$ in the numerator of
Eq.~(\ref{eq:calc}) and $c/N$ in the denominator to obtain:
\begin{equation}
N =  {{\alpha-1} \over { \alpha}} \left[ { - 1}\over
 {{ \left(c \over N \right)}^{\alpha} } \right].
\label{eq:calc2}
\end{equation}
By taking the logarithm on both sides and neglecting 
terms of order $1$ we obtain
\begin{equation}
\alpha \cong { \ln N \over {\ln (N/c)} }.
\label{eq:c->0}
\end{equation}
Note that even for systems in which the lower bound 
(which is due to some microscopic discretization) 
given by $c$, is orders of magnitude smaller than $1/N$,
the resulting
$\alpha$ may differ significantly from the free 
multiplicative random walk result 
$\alpha = 0$.
Since $c$ enters in the formula 
(\ref{eq:c->0}) 
for $\alpha$ through its logarithm,
the system gives away information on its 
microscopic scale cut-off  $c$ through the exponent
$\alpha$ 
of its macroscopic power law behavior. 

One should emphasize that in the region where 
$\alpha < 1$ the average $\bar w$ of 
the distribution 
$p(w)$ in 
Eq.~(\ref{eq:mean}) 
is not well defined and in fact one expects in the actual 
runs very wide macroscopic fluctuations
of this mean. 
These fluctuations are however never infinite because according 
to the formulae above,
as one increases the size of the system $N$, the region 
along the $c$ axis
where $\alpha < 1 $ shrinks to 0.
For $ 1 < \alpha < 2$ it is only the standard 
deviation of the distribution 
$p (w)$ which is formally divergent. 
This gives rise in the actual  
computer simulations to
wide fluctuations of the individual values of  $w_i$.
However, this divergence is  
kept in check too by the fact that no $w_i $ can 
possibly exceed 
$N \cdot \bar w$, 
namely
$ p (N \cdot {\bar w} ) = 0$.
This amounts to a {\it truncation from above} of  the power law 
Eq.~(\ref{eq:power}). 

\section{Numerical Simulations and Results}
\label{simulations}

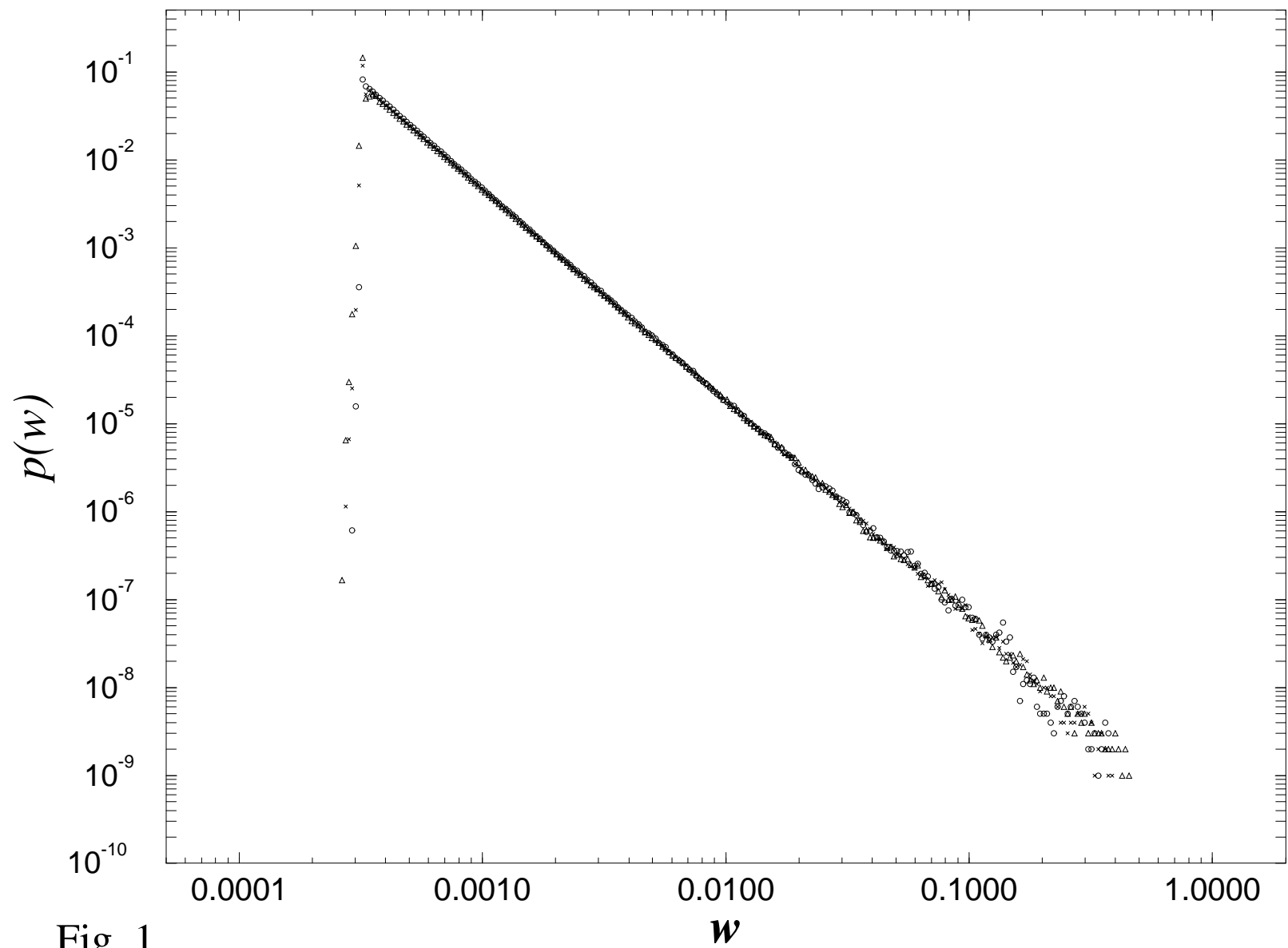
\begin{figure}
\caption{The distribution of the variables $w_i$, $i=1,\dots,N$
for $N=1000$ 
obtained from a numerical simulation of the model given by 
Eqs. 
(\ref{eq:mult}) 
and
(\ref{eq:subsidize}) 
with the lower cutoff 
$c=0.3$ and 
$ \Pi(\lambda)$ 
uniformly distributed in the range $0.9 < \lambda < 1.1$.
The distribution (presented here on a $\log-\log$ scale) exhibits
 a power law behavior described by
$p(w) \sim w^{-1-\alpha}$, where $\alpha = 1.4$.}
\label{fig:pareto}
\end{figure}

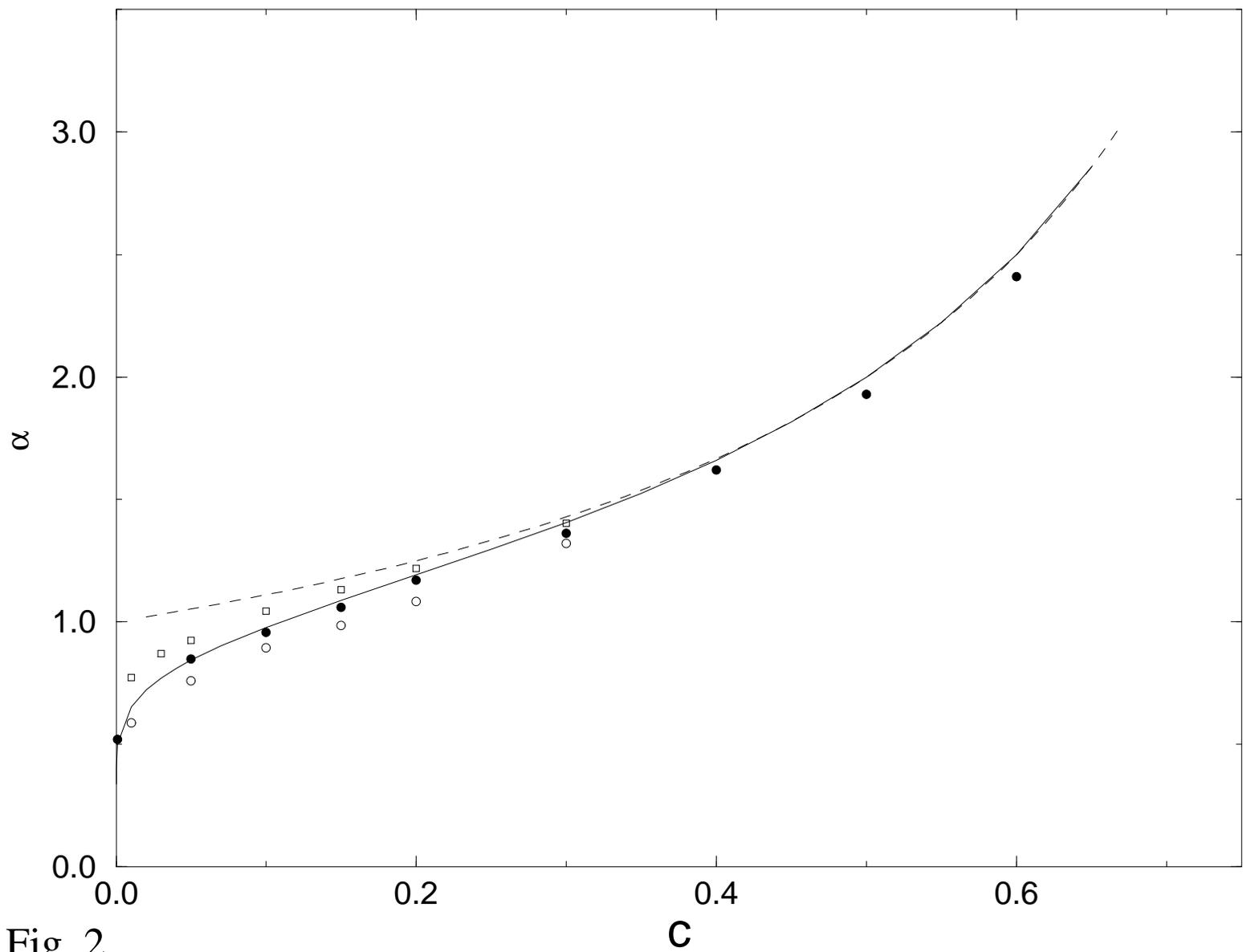
\begin{figure}
\caption{The exponent $\alpha$ of the power-law 
distribution of the variables
$w_i$, $i=1,\dots,N$
as a function of the lower cutoff $c$.
The data were obtained
 from the simulations
of the multiplicative stochastic process of
Eqs.~(\ref{eq:mult})
and
(\ref{eq:subsidize}) 
with
$N=100$ (empty dots),
$1000$ (full dots) 
and 
$5000$ (squares).
The theoretical 
prediction of
Eq.~(\ref{eq:calc})
is shown for $N=1000$ (solid line) and is in
 excellent
agreement with the numerical values
for all values of $c$.
The approximate 
expression of
Eq.~(\ref{eq:temp})
is also shown (dashed line).}
\label{fig:alpha}
\end{figure}

Numerical simulations 
of the stochastic multiplicative process 
described by Eqs.
(\ref{eq:mult})
and
(\ref{eq:subsidize}), 
confirm the validity of
Eq.~(\ref{eq:power}) 
for a wide range of lower bounds $c$. 
It appears that the exponent
$\alpha$ is largely independent of the shape of the probability distribution
$\Pi(\lambda)$. 
Fig. \ref{fig:pareto}
shows the distribution of $w_i$, $i=1,\dots,N$,
obtained for $N=1000$, $c=0.3$, and $\lambda$ 
uniformly distributed in the range
$0.9 \leq \lambda \leq 1.1$.
A power law distribution is found for a range
of three 
decades between $w_{min}=0.0003$
and $w_{max}=0.3$.
The slope of the best linear fit
 within this range is given by $\alpha=1.4$,
in agreement with Eqs. (\ref{eq:calc})
and
(\ref{eq:temp}).
On the horizontal axis of this graph 
the sum of all $w_i$'s
is normalized to 1 and therefore
$\bar w = 0.001$.
The exponent $\alpha$ as a function of the lower
cutoff $c$ is shown in Fig. \ref{fig:alpha}. 
Numerical results are presented 
for $N=100$ (empty dots),
$1000$ (full dots) and $5000$ (squares).
The prediction of 
Eq.~(\ref{eq:calc})
is shown for $N=1000$ (solid line), 
which is in good agreement with the numerical 
results for all values of $c$.
The approximate expression  
Eq.~(\ref{eq:temp})
is also shown (dashed line).
It is observed that for $N = 1000$ 
this approximation gradually starts to hold 
as $c$ is increased beyond
$1/ \ln(1000)$,
in agreement with
the theoretical analysis. 
In general, for a given $N$,
$\alpha$ 
is monotonically increasing as a function of
$c$, starting from
$\alpha = 0$ (which corresponds to $1/w$ distribution)
at $c=0$, where the $w_i$'s are uncoupled.
It is also observed that 
as $N$ is raised, the value of $\alpha$ which
corresponds to a given $c$ increases monotonically.  
As a result, the range of
validity $1/ \ln N \ll c < 1$  of the approximation 
Eq.~(\ref{eq:temp}) 
is extended and the knee 
adjacent to $c=0$ sharpens and becomes
a discontinuity for $N \rightarrow \infty$.
The range $0 \ll c < 1/N$ in which the approximation 
of
Eq.~(\ref{eq:c->0}) 
is valid,
shrinks correspondingly.

Let us turn now to the dynamics of the system as a whole.
The dynamics of the system involves, according to
Eq.~(\ref{eq:mult}), 
a generalized random walk with step sizes distributed according to 
Eq.~(\ref{eq:power}).
Therefore, the stochastic fluctuations of $\bar w (t)$ 
after $\tau$ time steps:
\begin{equation}r(\tau) = {{{\bar w(t+\tau)} - {\bar w(t)}} \over
{\bar w(t)}}\label{eq:return}
\end{equation}
\noindent
are governed 
\cite{Mantegna94} by a truncated L\'evy distribution $L_{\alpha}(r)$.

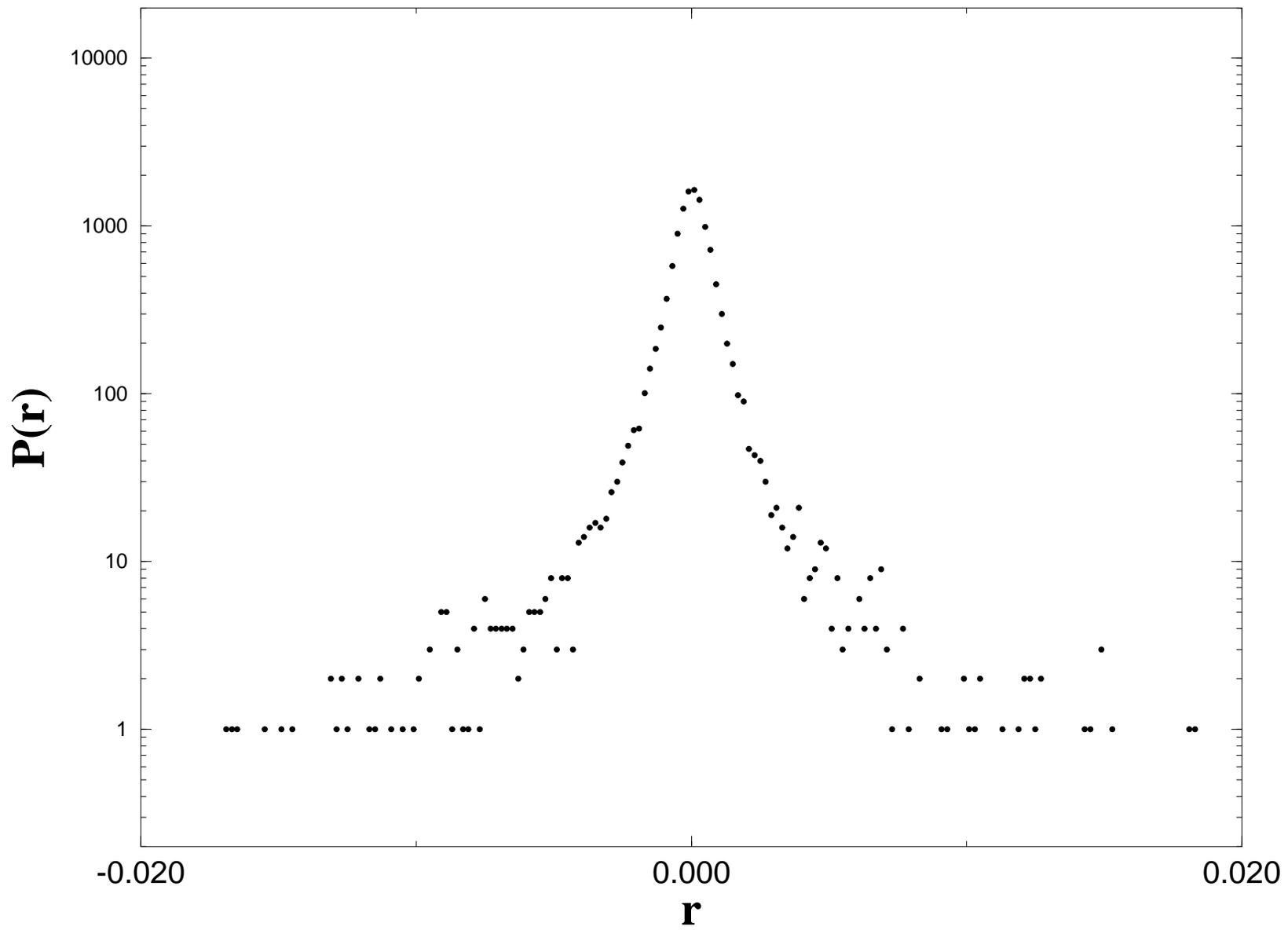
\begin{figure}
\caption{The distribution of the variations of $\bar w$ after 
$\tau$ steps
$r(\tau) = [{\bar w(t+\tau)} - {\bar w(t)}] /{\bar w(t)}$
where
$\tau=50$, for the same parameters as in Fig. \ref{fig:pareto}.
This distribution has a L\'evy-stable shape with index
$\alpha =1.4$.}
\label{fig:levy}
\end{figure}

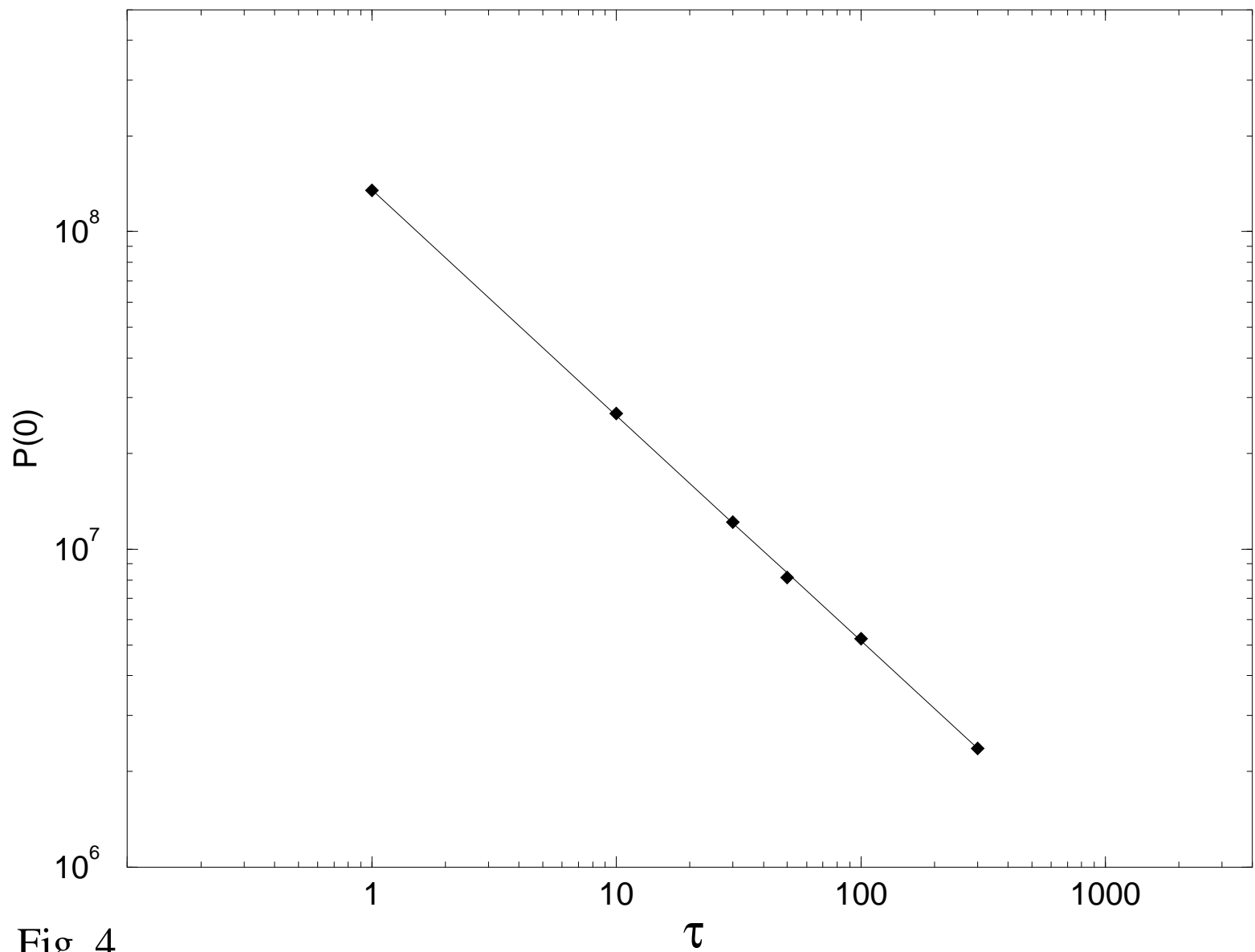
\begin{figure}
\caption{The scaling with 
$\tau$ of the probability that
$r(\tau) = [{\bar w(t+\tau)} - {\bar w(t)}] /{\bar w(t)}$ is 0.
The parameters of the process are the same as in Figs. 
\ref{fig:pareto} 
and 
\ref{fig:levy}.
The slope of the straight line on the logarithmic scale is
$0.71$ which corresponds to a L\'evy-stable 
process with $\alpha = 1/0.71 = 1.4$.}
\label{fig:levyzero}
\end{figure}

In  Fig. \ref{fig:levy}
we show the distribution
of the stochastic fluctuations
$r(\tau)$ for $\tau = 50$,
which is given by a (truncated)
L\'evy distribution 
$L_{\alpha}(r)$.
According to
Ref.
\cite{Mantegna95},
the peak of the (truncated) L\'evy-stable 
distribution scales with $\tau$  as
\begin{equation}
L_{\alpha}(r=0) \sim \tau^{-1/\alpha}
\label{eq:fluct}
\end{equation}
\noindent
where $\alpha$ is the index of the L\'evy  distribution.
In 
Fig. \ref{fig:levyzero}
we show the height of the peak  P$(r=0)$ of 
Fig. \ref{fig:levy} 
as a function of  $\tau$.
It is found that the slope of the fit in 
Fig. \ref{fig:levyzero}
is $-0.71$, which 
following  the scaling relation 
(\ref{eq:fluct}) 
means that the index of the
L\'evy distribution in 
Fig. \ref{fig:levy} 
is $\alpha= -1/(-0.71) = 1.4$.
These results were obtained for the same parameters 
which gave rise to the power law 
distribution with $\alpha=1.4$ in 
Fig. \ref{fig:pareto}.
Thus, the prediction that the fluctuations of 
$\bar w$ in Fig. 
\ref{fig:levy} 
follow a
(truncated) L\'evy-stable distribution with an 
index $\alpha$ which equals the exponent $\alpha$
of the power-law distribution in 
Fig. \ref{fig:pareto}, 
is confirmed.

\section{Discussion}

The model considered in this paper may be relevant 
to a variety of empirical systems 
in the physical, biological and social sciences
which can be described by a set of interacting
dynamic variables which follow a stochastic 
multiplicative dynamics.
Such dynamical processes may play a role in the formation of the
mass distribution in the universe where clusters of galaxies
accumulate and eventually form super-clusters.
In a different context, the growth of cities is 
basically a multiplicative process governed by the reproduction rate
of the local population in addition to mobility between cities.

Enhanced diffusion processes, which can be described by the
L\'evy-stable distribution have been observed in a variety of
nonlinear dynamical systems
\cite{Geisel1987,Zumofen1994}.
Unlike the stochastic model studied here, these systems are
governed by deterministic rules. They exhibit intermittent
chaotic motion which gives rise to enhanced diffusion.

In population dynamics, the number of individuals in 
each specie varies stochastically from one season to the next
with a multiplicative factor which depends on the local conditions.
The lower bound may represent the minimal number of individuals
required for the species to survive in the given environment. 
In this case the number of species
may not be strictly a constant, but species that are wiped out may be
replaced by others which invade their area.
In this context it was found
that the number of species of a given size
often
follows a decreasing power-law distribution as a function of their size
(see e.g. Ref. \cite{Morse85}).

In the economic context of
a stock-market system 
the dynamic variables 
$w_i$, $i=1,\dots,N$ 
may represent the wealth of individual investors.
In this case the dynamics 
represents the increase (or decrease) by a random factor 
$\lambda(t)$ of the wealth
$w_i$ of the investor $i$ between times $t$ and $t+1$. 
The lower bound may represent a minimal wealth 
required in order to participate in stock market trading.
In a more general economic model, 
this lower bound may be related to 
a basket of basic publicly funded services which every 
individual receives.
In another possible interpretation, the 
$w_i$'s
represent the capitalization (total market value) 
of the firm $i$,
which may increase (or decrease) 
by a factor
$\lambda(t)$
at each time step.
In this case the lower bound may represent
the minimal requirements for a company stock to be publicly traded.

Studies of the
distribution of wealth in the 
general population revealed a power-law
behavior 
(see e.g. Ref. 
\cite{Atkinson78}).
More recently
it was shown 
\cite{Levy97}
that the 
distribution of individual wealth of the 400 richest 
people in the United States 
(Forbes 400) 
corresponds to a power law with $\alpha= 1.36$
[more precisely 
$W(n) = C \cdot n^{-1/\alpha}$ 
where $W(n)$ is the wealth of the
$n$-th richest person on the list]. 
Recent analysis of stock market returns, 
measured over many years 
found a truncated L\'evy distribution 
$L_{\alpha}(r)$ with the index $\alpha = 1.4$
for an extended (but finite) range of returns $r$
\cite{Mantegna96}.
These results indicate that the property observed in our model,
namely that the same value of the index $\alpha$ appears both in
the power law distribution and in the
L\'evy-stable distribution of the fluctuations 
may be of relevance in the
economic context. 
To further explore this possibility 
it would be interesting to 
examine whether
the distribution of
total market values of
companies in the stock market exhibits a power law behavior
of the form
(\ref{eq:power})
with $\alpha = 1.4$.

\section{Summary}

We have studied a generic
model of stochastic auto-catalytic dynamics of 
many degrees of freedom using computer
simulations. 
The model consists of dynamic variables $w_i$, 
$i=1,\dots,N$ which are 
updated randomly one at a time through an 
autocatalytic process at the individual
level. 
In addition, the variables are coupled 
through a lower bound constraint which 
enhances the variables which fall below a 
fraction of the global average.
The model may describe a large variety of 
systems such as stock markets and
city populations. The distribution $p(w,t)$
of  the system components $w_i$
turns out to fulfill a power law distribution 
of the form 
$p(w,t)  \sim w^{-1-\alpha}$.
In the limit $N= \infty$, $c \rightarrow 0$ one obtains the
case often encountered in nature: $\alpha \approx 1$.
The average ${ \bar w (t)} $
exhibits intermittent fluctuations following a
L\'evy-stable distribution with the same index $\alpha$. 
This relation between 
the distribution of system components and the 
temporal fluctuations of their average 
may be relevant to
a variety of empirical systems.
For example, it may provide a connection between the distribution of 
wealth/capitalization in a stock market and the 
distribution of the index fluctuations.

\onecolumn

\end{document}